\documentstyle[12pt]{article}
\input math_macros.tex

\begin{document}
\begin{titlepage}
\begin{center}

January 23, 1998     \hfill    LBNL-41288 \\

\vskip .5in

{\large \bf Comment on Unruh's Paper}
\footnote{This work was supported by the Director, Office of Energy 
Research, Office of High Energy and Nuclear Physics, Division of High 
Energy Physics of the U.S. Department of Energy under Contract 
DE-AC03-76SF00098.}
\vskip .50in
Henry P. Stapp\\
{\em Lawrence Berkeley National Laboratory\\
      University of California\\
    Berkeley, California 94720}
\end{center}

\vskip .5in

\begin{abstract}
William Unruh has commented on my recent proof of the incompatibility
of certain predictions of quantum theory with the idea that no influence of 
any kind can act backward in time in any Lorentz frame. He argues that my 
proof contains, contrary to my explicit claim, a tacit reality assumption 
that conflicts with the ideas quantum theory. I show here that Unruh's
argument fails to draw a necessary distinction between statements that make 
an assertion about actual values and statements that make an assertion about 
correlations between possible values, and that consequently his argument
fails demonstrate the occurrence in my proof of any reality assumption 
that is alien to the principles of quantum theory.

\end{abstract}
\medskip
\end{titlepage}

\renewcommand{\thepage}{\roman{page}}
\setcounter{page}{2}
\mbox{ }

\vskip 1in

\begin{center}
{\bf Disclaimer}
\end{center}

\vskip .2in

\begin{scriptsize}
\begin{quotation}
This document was prepared as an account of work sponsored by the United
States Government. While this document is believed to contain correct 
 information, neither the United States Government nor any agency
thereof, nor The Regents of the University of California, nor any of their
employees, makes any warranty, express or implied, or assumes any legal
liability or responsibility for the accuracy, completeness, or usefulneents
that its use would not infringe privately owned rights.  Reference herein
to any specific commercial products process, or service by its trade name,
trademark, manufacturer, or otherwise, does not necessarily constitute or
imply its endorsement, recommendation, or favoring by the United States
Government or any agency thereof, or The Regents of the University of
California.  The views and opinions of authors expressed herein do not
necessarily state or reflect those of the United States Government or any
agency thereof or The Regents of the University of California and shall
not be used for advertising or product endorsement purposes.
\end{quotation}
\end{scriptsize}

\vskip 2in

\begin{center}
\begin{small}
{\it Lawrence Berkeley Laboratory is an equal opportunity employer.}
\end{small}
\end{center}

\newpage
\renewcommand{\thepage}{\arabic{page}}
\setcounter{page}{1}

Unruh's essential point$^1$ in expressed in his two consecutive paragraphs 
that read as follows:

``However, great care is required in such counter-factual statements that 
one does not import into the statements a notion of reality. In particular,
the truth of the statement made about system A which relies on measurement 
made on system B and on the correlations which have been established between
A and B in the state of the joint system is entirely dependent on the truth 
of the actual measurement which has been made on system B. To divorce them is
to effectively claim that the statement made about A can have a value in and 
of itself, and independent of measurements which have been made on A. This 
notion is equivalent to asserting the reality of the statement about A 
independent of measurements, a position contradicted by quantum mechanics. 

``Thus in the above system, measuring $\sigma _z$ on particle one can lead 
to one assigning a value + to $\sigma _x$ of particle 2, even if that 
attribute was not measured, due to the correlation between the two particles. 
However, that value for $\sigma _{2x}$ is entirely dependent upon the fact
that $\sigma _x$ was actually measured on particle 1.  In particular, 
causality cannot be used to argue that the inferred fact of the value of
$\sigma _{2x}$ must be independent of what was measured at particle 1.
Such an extension of the concept of locality is inapplicable to a quantum
system, and is certainly not necessary to capture the concept of locality.'' 

The first paragraph already contains some linguistic irregularities that are
harbingers of trouble ahead. In order to be relevant to the case at hand 
[see below] one should take ``relies on measurement made on system B'' to
mean ``relies upon the fact that a certain specified measurement is made on
system B''. And  ``the truth of the actual measurement which has been 
made on system B'' should mean ``truth of the statement that the
specified measurement has been made on system B.'' Also, ``the 
statement made about A can have a value in and of itself,'' should mean 
``the statement made about A can have a truth value in and of itself.''
And the ``reality of the statement about A'' should be ``the truth of the 
statement about A''.

The linguistic irregularities in Unruh's original wording appear to stem 
from an effort to make one wording cover two very different situations. The 
first case is the familiar one in which certain outcomes, hence {\it values}, 
in the two different systems are correlated, so that the {\it value} obtained
in one region fixes the {\it value} that would be found in the other system 
if the appropriate measurement were actually performed there. The second case 
is the one that occurs in my proof. There the {\it truth of a statement
about  correlations between possible outcomes in one system} is dependent 
solely upon {\it which measurement is performed} on a faraway system 
at a later time.  Unruh's argument involves confounding these two very 
different cases.

In the first case the regular wording of the paragraph would be this:

``However, great care is required in such counter-factual statements that 
one does not import into the statements a notion of reality. In particular,
the truth of the statement made about {\it a value in} A which relies on 
{\it a value obtained from a} measurement made on system B and on the 
correlations which have been established between {\it values in} A and B in 
the state of the joint system is entirely dependent on the fact that the
actual measurement has been made on system B. To divorce them is to 
effectively claim that A can have {\it a value} in and of itself, and 
independent of measurements which have been made on A. This notion is 
equivalent to asserting the reality of the {\it value in} A independent of 
measurements, a position contradicted by quantum mechanics.''

This is the normal sort of quantum reasoning. 

But to make the statement cover the relevant step in my proof the statement
should read:

``However, great care is required in such counter-factual statements that 
one does not import into the statements a notion of reality. In particular,
the truth of the statement made about {\it correlations in} system A which  
relies upon the fact that a certain specified measurement is made on system B, 
and on the correlations which have been established between A and B in the 
state of the joint system is entirely dependent on the truth of the statement 
that the specified measurement has been made on system B.  To divorce them is
to effectively claim that the statement made about {\it correlations in} A can 
have a truth value in and of itself, and independent of measurements which 
have been made on A. This notion is equivalent to asserting the truth of the 
statement about A independent of measurements, a position contradicted by 
quantum mechanics.'' 

Now everything again fits together linguistically, and fits the context 
of my proof. But the truth of the final assertion ``a position contradicted 
by quantum mechanics'' is not so obvious: why cannot the truth of a statement 
about A be independent of which measurement is performed faraway on system B 
at a later time, when the statement in question merely asserts the existence 
of a certain {\it correlation} between the outcomes of possible measurements 
in A, not the existence of any actual {\it values} in A.

Indeed, the theory of relativity suggests that the truth of such a statement
about system A should not depend upon which measurement on the faraway system 
B is freely chosen at the later time.

Let us examine the matter more closely.

The step in my argument that is under attack here is the step from Line 5
to Line 6. Line 5 has been proved by using three assumptions: 1) the choices
made by the experimenters can be treated as free (unconstrained) variable;
2) a locality assumption LOC1 that expresses the idea that the macroscopic
outcomes that appear in one region are independent of what a faraway 
experimenter chooses to do at some later time; and 3) the assumption that 
if any of the experiments under consideration here is performed then nature 
will produce an outcome that is in accord with the predictions of quantum 
theory. 

The result stated in Line 5 is not being challenged. It reads:

Line 5: ``If L2 is performed then statement SR is true.''

Here SR is the statement:

SR: ``If R2 is performed in region R, and the outcome there is +, then
if R1, instead of R2, had been performed in region R the outcome there
would have been --.''

The region L is supposed to later in time than region R.

Then my proposed locality assumption LOC2 is:

LOC2: ``If SR is proved true under the condition that L2 is freely chosen
in region L then SR must be true also under the condition that L1 is 
chosen there.''

That is, LOC2 asserts that ``If Line 5 is true then Line 6 is true'',
where Line 6 is:

Line 6: ``If L1 is performed then statement SR is true.''

The three essential points are: 

1) LOC2 is an assertion about the {\it truth} of the statement SR, 
which is a conditional statement whose premise is that R2 is 
performed and the outcome is +; it it not an assertion that some 
outcome, or {\it value}, is fixed in nature.

2) No outcome of the experiment in region L is mentioned; the only conditions
pertaining to outcomes ({\it i.e., values}) occur within the statement SR,
which is an assertion of the existence of a {\it correlation}
between outcomes of two alternative possible measurements in region R.

3) I claim that LOC2 is a locality condition: it expresses the condition
that the property described by SR, which asserts the existence of a certain
correlation between outcomes of possible measurements in region R, cannot
depend upon which experiment is freely chosen later in region L
 
Unruh makes his contradictory claim in the second paragraph quoted above: 

``causality cannot be used to argue that the inferred fact of the value of 
[the unmeasured quantity] must be independent of what was measured [faraway].'' 

However, this wording contains an essential ambiguity: it does not distinguish
between two cases that are structurally very different, due to the different
character of the inference that is involved. Unruh justifies this claim
by considering one of these cases, but applies it to the other case.

The case in which the claim is justified is the one in which one infers
the existence of an actual value in one system from the value measured on the 
faraway system. That is the usual familiar case.

But LOC2 does not claimed that some quantity associated with an unperformed 
experiment has an actual value.  LOC2 is a different kind of claim: it asserts 
the {\it truth} of a statement that has a condition and a counterfactual 
consequence. It does not assert the factual {\it value} of an unmeasured 
quantity that can be direcly inferred from the knowledge of the outcome of 
an experiment performed faraway.

This may seem like a distinction too fine to matter. But it does matter.
For no reality condition alien or contrary to quantum thinking is required
to give a meaning to LOC2 that: 1) is completely and naturally specified by 
the words in the statement that define it; and 2) entails that a failure of 
LOC2 would mean the existence of some sort of action or influence backward 
in time. 

In the way of speaking used by Dirac, Heisenberg, most quantum physicists, 
and occasionally even Bohr, if one of these experiments is performed then
nature produces (or chooses) an outcome. Statement SR, expressed in this 
language, asserts that:

SR': ``If under the condition that R2 is performed in region R nature 
produces outcome + in region R,  then if R1 had been performed there, 
instead of R2, nature would have produced there the outcome --.''

The concept of ``what would have happened'' if the free choice in region R
had gone the other way was introduced in LOC1, and it arises not from any
assumption of determinism, but as an expression of the idea that what has 
already happened at the macroscopic level, and been observed by a human 
observers, and been recorded in some memory device, cannot depend upon 
a later free choice by a faraway experimenter as to which experiment he will 
perform at some still later time.

Statement SR' has a meaning independently of whether it can be proved
under certain conditions on the predictions of quantum theory. Generally
this condition SR' is not true, and can be proved false independently of 
which experiment is performed in region L. But under the special 
circumstances of the Hardy experiment the property specified in SR' can be 
proved to hold under the condition that L2 be performed, without any condition 
on what the outcome of that experiment L2 is.  

One cannot prove in the same way that the property described in SR' 
would necessarily hold also if the later free choice had gone the other
way. But I believe that if SR' necessarily holds if the later free 
choice is L2, but would fail to hold if that later free choice had gone the 
other way, then there must be {\it some sort} of backward-in-time effect of 
that later free choice. 

This argument does not involve any idea of reality alien to quantum theory.
It does not require, as Unruh appears to claim, and it does not 
even suggest, that nature has a hidden variable that specifies what the 
outcome of R1 would be, independently of any outcome of any experiment. 
SR' asserts rather that there is a {\it correlation} between the outcomes 
of different alternative possible measurements. Correlations between 
outcomes of possible measurements are a standard feature of quantum theory, 
and the extension of such correlations to alternative possible measurements 
follows in the present case from the locality condition LOC1, which asserts 
a trivial correlation (nondependence) of the outcomes of a certain measurement
on a free choice between two (faraway) alternative possible experiments.
The correlation specified SR' follows, in case L2 is performed, from 
three correlations linked in tandem.

This detailed examination shows that Unruh's argument does not identify in 
my proof any tacit or hidden reality assumption that is alien to the 
principles of quantum theory.

\noindent {\bf References}

1. W. Unruh, {\bf Is Quantum Mechanics Non-Local?} \\
http://quantum-ph/9710032@xxx.lanl.gov

2. Henry P. Stapp, {\it Nonlocal character of Quantum theory,}
Amer. J. Phys. {\bf 65}, 300-304 (1997)\\
http://www-physics.lbl.gov/\~{}stapp/stappfiles.html
\end{document}